\documentclass[showpacs,preprint,amsmath,amssymb]{revtex4}
\usepackage{graphicx}
\usepackage{psfig}
\usepackage{graphics}
\usepackage{amsmath}
\usepackage{dcolumn}
\usepackage{bm}

\begin{document}

\title{The self-energy of a charged particle in the presence of a
topological defect distribution}

\author{A. M. de M. Carvalho and Fernando Moraes}
\affiliation{ Laborat\'orio de F\'{\i}sica Te\'orica e Computacional,\\
Departamento de F\'{\i}sica,\\ Universidade Federal de Pernambuco,
50670-901, Recife, PE, Brazil}

\author{Claudio Furtado}
\affiliation{Departamento de F\'{\i}sica, CCEN,  Universidade Federal 
da Para\'{\i}ba, Cidade Universit\'{a}ria, 58051-970 Jo\~ao Pessoa, PB,
Brazil}

\begin{abstract}
In this work we study a charged particle in the presence of both a
continuous distribution of disclinations and a continuous
distribution of edge dislocations in the framework of the
geometrical theory of defects. We obtain the self-energy for a
single charge both in the internal and external regions of either
distribution. For both distributions the result outside the defect
distribution is the self-energy that a single charge experiments 
in the presence of a single defect. 
\end{abstract}
\maketitle

\section{Introduction}

In recent years many condensed 
matter systems have been identified as analogue models of gravity:
superfluid helium~\cite{boo,prd:unruh}, Bose-Einstein 
condensates~\cite{prl:gar,pra:gar,cqg:visser}, as well as non-linear 
electrodynamics~\cite{novello} are examples of such systems. The geometrical 
description of defects in elastic media provides another important
analogue 
model: elastic solids with defects can be mapped into three-dimensional 
gravity with torsion~\cite{katanaev}.In this model, the disclination in 
the elastic solid corresponds to the space section of the cosmic string. 
Electronic properties of solids with  defects have been traditionally 
described by the deformation  potential theory~\cite{pr:bs}, which adds 
to the electronic Hamiltonian an effective interaction potential due to 
the presence of the defect in the medium. This potential is obtained
from 
the elastic deformation of the medium caused by the defect.
Alternatively, the analogue model developed by Katanaev and 
Volovich~\cite{katanaev} gives a simple geometric description for the
elastic 
medium with defects. This framework 
has been extensively used to study effects like bound
states~\cite{furtado}, 
Aharonov-Bohm phases~\cite{furtado2,furtado3}, etc. caused by
topological 
defects in solids. These results are in agreement we the standard 
deformation potential theory, but with clear geometric interpretation
and easy computation. In this formalism the boundary conditions imposed by the 
defects in the elastic continuum are accounted for by a non-Euclidian
metric. In general the defects correspond to singular curvature 
or torsion (or both) along the defect line~\cite{katanaev}.  
The advantage of this geometric description of defects in solids 
is twofold. First, in contrast to the ordinary elasticity theory 
this approach provide an adequate language for continuous distribution 
of defects~\cite{katanaev2}. Second, the mighty
mathematical machinery of differential geometry clarifies and
simplifies the calculation.

The deformation potential turns out to be equivalent to the
self-interaction of a point particle in the background space of a
continuous distribution of topological defects. The self-interaction is 
a particular case of the well known fact~\cite{dewitt,vilenkin} that a 
point charge in the static gravitational field experiences an
electrostatic 
force due to the change in the boundary conditions its electric field is 
submitted to. In other words, the change in the geometry of the
space-time 
which is associated with the gravitational field causes a deformation of
the
electromagnetic fields lines that induces a self-force on the
charge. 

The interest on self-forces on charge distributions due to topological
defects 
started when Linet~\cite{linet} and Smith~\cite{smith} found the
electrical
self-force on a point charge in the presence of a cosmic string. A
variety 
of studies involving self-forces on charge distributions due to
topological 
defects have appeared in the recent 
literature~\cite{souradeep,cqg:furtado,guimaraes,mello1,mello2,carvalho,
azevedo,lorenci}.  All these works refer to linelike, i.e. extremely
thin, 
line defects. More realistic, finite thickness defects have been studied 
under this point of view only  recently: the electrostatic 
self-interaction acting on a single charge and on a linear charge 
distribution in the presence of the Gott-Hiscock cosmic 
string~\cite{val,fer}, respectively. In this article, we are interested 
in generalising references~\cite{mello1},the study of the electric  
self-force on a linear charge distribution
parallel to a cosmic string, and ~\cite{carvalho}, the case of a linear
charge 
distribution in the presence of an edge dislocation, to a continous 
distribution of line defects, in order to take into account the finite 
thickness of the defects. Therefore, we study the classical effects  a 
continuous distribution of line defects causes in the
electric field of a charged particle, both in the case of the
disclination 
distribution and for the edge dislocation distribution. This paper is  
organized as follows, in section II we present the geometric description
of 
the density of line defects, for both cases. In section III, we find the
two-dimensional Green functions for a charged particle in the
presence of a defect distribution. In section IV we evaluate the 
self-energy and the self-force for a linear charge distribution in the 
presence either of a distribution of disclinations or a distribution of 
dislocations. Finally,  section V is dedicated to 
the conclusions.

\section{The defect distribution }

In the geometric theory of defects in solids the techniques of differential geometry 
are used to describe the strain and
stress fields induced by the defect in the elastic medium. 
All information on these fields is  contained in the
geometric quantities (metric, curvature tensor, etc.) that describe the 
elastic medium with defects.  Alternative approaches for geometric descriptions of 
defects have been presented by several authors~\cite{klei,hol,put,kol,baush}.
The geometric theory of defects in solids represents the elastic
deformations in the continuum medium by a non-euclidean metric. This theory can be used to
describe a medium either with a single line defect or a density of
defects. In this section we use this theory to analyze
the density of wedge disclinations and edge dislocations. In a recent paper 
Katanaev~\cite{katanaev3}  demonstrated that solutions  of the geometric 
theory of defects yields the solutions of the same  problems in  the 
nonlinear elasticity theory. Morever, the metric that describes a defect solution 
is an exact solution  and  the linear elasticity is obtained 
taking the appropriate limits.

We consider a distribution of linear defects in a finite region of space.
We suppose that all defects are ligned with the
$z$-axis and also that the distribution is circularly symmetric.
We apply the method developed by Katanaev and Volovich~\cite{katanaev2}
to handle with a continuous distributions of defects.

\subsection{The disclination distribution}

In this subsection we analyze the density of disclinations.
Disclination in solids is  the  analogue of a single particle
in $2+1$-dimensional gravitation \cite{katanaev}. In this formalism the defect
formation can be viewed as a ``cut and glue'' process, known in 
literature as the Volterra process. The disclination is obtained by
either removing or inserting material in the medium.
We consider a very concentrated number of disclinations in a circular 
region of space. 

Consider a circularly symmetric distribution of disclinations on a
disk of radius $R_{0}$, which acts like a source of the distortion field.
This density can be written as ~\cite{katanaev2}:
\begin{equation}
\lambda (r)=\left\{
\begin{array}{c}
q,\qquad r\leq R_{0} \\
0,\qquad r > R_{0}
\end{array}
\right. .  \label{densidade}
\end{equation}
The normalized deficit/excess angle is given by the integral
\begin{equation}
\Theta  =\frac{1}{2\pi }\int \lambda (r)dA =\frac{1}{2}qR_{0}^{2}.
\label{deficit}
\end{equation}
Since our problem is effectively two-dimensional, the space around the density of 
defects is described by a conformal line element
\begin{equation}
ds^{2}=\exp (-\Omega )\left( dr^{2}+r^{2}d\theta ^{2}\right),
\label{mconforme}
\end{equation}
where $\Omega$ is the conformal factor. This
metric must be a solution of Einstein equation
\begin{equation}
R_{\alpha \beta }-\frac{1}{2}g_{\alpha \beta }R
=-8\pi GT_{\alpha \beta },  \label{einstein}
\end{equation}
where $R_{\alpha \beta }$ is the Ricci tensor, $R$ the Ricci scalar,
$G$ a ``gravitational constant''and $T_{\alpha \beta }$ is the 
energy-momentum tensor. In fact, $G,$ in condensed matter is a 
constant associated to the continuum elastic medium, and the 
tensor $T_{\alpha \beta }$ is the
source the strain and stress fields, ${\it i.e.}$ the density of defects itself. 
Then density of deficit angles is described by
the $T_{zz}$ component of the energy-momentum tensor. With
knowledgment of the metric we can easily obtain the Ricci scalar
\begin{equation}
R=-e^{\Omega }\Delta \Omega .  \label{escalar-curv}
\end{equation}
and then obtain the differential equation for the conformal factor,
\begin{equation}
\Delta \Omega =2\lambda .  \label{eq-dif-conf}
\end{equation}
The solution of this equation is
\begin{equation}
\Omega \left( r\right) =\left\{
\begin{tabular}{ll}
$a+b\ln r+\frac{1}{2}qr^{2},$ & $r\leq R_{0}$ \\
$c\ln r+d,$ & $r > R_{0}.$%
\end{tabular}
\right. 
\end{equation}
This solution and its derivative  satisfy the following boundary
condition on the interface:
\begin{eqnarray}
\left\{
\begin{array}{c}
\Omega _{in} \left( R_{0}\right) =\Omega _{out} \left(
R_{0}\right)  \\
\Omega _{in}^{\prime }\left( R_{0}\right) =\Omega _{out%
}^{\prime }\left( R_{0}\right).
\end{array}
\right.
\end{eqnarray}
Applying this boundary condition it is obtained
\begin{equation}
\Omega \left( r\right) =\left\{
\begin{tabular}{ll}
$qR_{0}^{2}\ln R_{0}-\frac{1}{2}qR_{0}^{2}+\frac{1}{2}qr^{2},$ &
$r\leq R_{0}$ \\
$qR_{0}^{2}\ln r,$ & $r > R_{0},$
\end{tabular}
\right. \label{cfactordisc}
\end{equation}
which is the result of Katanaev and
Volovich ~\cite{katanaev2}. Notice that, far away from the distribution, 
the metric corresponding to the density of disclinations is
equivalent
to that of a single disclination.

\subsection{The edge dislocation distribution }

Consider now a continuous distribution of edge dislocations,
characterized by the density of the normalized Burger's vectors,
$\vec{\beta}(\vec{r})=(\beta_{x},0)$, where $\beta_{x}$ is given
by~\cite{katanaev2}
\begin{equation}
\beta_{x}=\left\{
\begin{array}{c}
\beta,\qquad r\leq R_{0} \\
0,\qquad r > R_{0}.
\end{array}
\right.   \label{densidade2}
\end{equation}
Notice that this density of defects behaves like a step function, so
the Burger's vector density can be written in terms of the Heaviside function
as $\vec{\beta}=\beta_{x}\Theta(R_{0}-r) \hat{\i}$. We also notice that
the
density is uniformly distributed on a disk of radius $R_{0}$.
For this distribution it is shown \cite{katanaev2} that
Einstein equation reduces to Poisson equation and,
furthermore, that the density of defects is written in terms of the
divergence
of $\vec{\beta}$ \cite{katanaev2} 
\begin{equation}
\triangle \Omega=2 (\vec{\nabla}\cdot  \vec{\beta}).
\end{equation}
Poisson equation is then rewritten as
\begin{eqnarray}
\triangle \Omega&=&2\beta \frac{\partial}{\partial x} \Theta(R_{0}-r) \\
  &=&-2\beta \cos \theta \delta(r-R_{0}).
\end{eqnarray}
The boundary condition now requires continuity of the
solution and  discontinuity of its derivative for the radial
part of the conformal factor, that is
\begin{eqnarray}
\left\{
\begin{array}{c}
\Omega _{in} \left( R_{0}\right) =\Omega _{out} \left(
R_{0}\right)  \\
\Omega _{out}^{\prime }\left( R_{0}\right) -\Omega _{in%
}^{\prime }\left( R_{0}\right)= -2\beta.
\end{array}
\right.
\end{eqnarray}
The following solution to the conformal factor is then obtained:
\begin{equation}
\Omega=\left\{
\begin{array}{c}
\beta r \cos\theta,\qquad r\leq R_{0} \\
\frac{\beta R_{0}^{2} \cos\theta}{r},\qquad r > R_{0}
\end{array}
\right. .  \label{cfactordisl}
\end{equation}
This result was first obtained by Katanaev and Volovich, like the one in the previous section, 
and it is included here for clarity reasons. Notice that, as in the density of disclinations 
case, far away from the distribution, 
the metric corresponding to the density of edge dislocations is
equivalent
to that of a single dislocation. 

\section{The two-dimensional Green function}

In this section we use the formalism developed by Garcia and Grats
~\cite{grats} to evaluate the self-force in an environment endowed 
with a conformal metric. Given
an arbitrary distribution of charge $\rho(\vec{x})$, the electrostatic
energy is given by \cite{jackson}
\begin{equation}
U_{ele}=-\frac{1}{2}\int d^{3}x\rho \left( \vec{x}\right) \Omega
\left(
\vec{x}%
\right) ,  \label{energia}
\end{equation}
where $\Omega(\vec{x})$ is the scalar potential. We are interested in
evaluating the self-energy and the self-force on a charged particle
due to the density of disclinations. We assume that the charge
distribution is punctual,
\begin{equation}
\rho(\vec{x})=Q\delta(\vec{x}-\vec{x'}),
\end{equation}
in this way we can write the electrostatic energy as
\begin{equation}
U_{ele}=-\frac{1}{2}\int\int
d^{2}xd^{2}x'\rho(\vec{x})G(\vec{x},\vec{x'})\rho(\vec{x'}).
\end{equation}
This leads to the following expression
\begin{equation}
U_{ele}=-\frac{Q^{2}}{2}G(\vec{x},\vec{x})|_{reg}, \label{uele}
\end{equation}
where, $G(\vec{x},\vec{x})|_{reg}$ is the regularized Green
function. The question now reduces to finding this function. 
In general, in a non-Euclidian metric, this is not an easy task. 
But the density of disclinations  shows an important characteristic: 
since the metric is effectively two-dimensional it is conformal to the Euclidean metric. 
This means that it can be written
as
\begin{equation}
g_{ab}=e^{-\Omega}\delta_{ab}.
\end{equation}
The Green function satisfies the two-dimensional Poisson equation
\begin{equation}
\Delta _EG(\vec{x}\,^{\prime },\vec{x})=\delta ^2(\vec{x}\,^{\prime
}-\vec{x}%
),
\end{equation}
where $\Delta _E$ is the two-dimensional Euclidean Laplacean. The
solution
of this equation is given by~\cite{jackson}
\begin{eqnarray}
G( \vec{x} \,^{\prime },\vec{x}) &=&\frac 1{4\pi }ln|\vec{x}\,^{\prime
}-\vec{x%
}|^2+\mbox{an arbitrary analytic function}  \nonumber \\
&&\mbox{function of}(\vec{x}\,^{\prime }-\vec{x}).  \label{green
geral}
\end{eqnarray}
The choice of this analytic function depends of the boundary
conditions (notice also that these functions must obey Laplace
equation).
It is very difficult to formulate these boundary conditions in a
general case. But we are effectively working in a two-dimensional
surface therefore it is natural that the field of a point source should tend to zero at
infinity. So, boundary conditions reduce the choice of the analytic
function to an arbitrary constant which can be taken equal to zero.
The Euclidean Green function (\ref{green geral}) reduces then to
\begin{equation}
G_0^{\left( 2\right) }(\vec{x}\,^{\prime },\vec{x})=\frac 1{4\pi
}ln|\vec{x}%
\,^{\prime }-\vec{x}|^2.
\end{equation}

The Euclidean Green function can be written in the following form
\begin{equation}
G_0^{\left( 2\right) }(\vec{x}\,^{\prime },\vec{x})=\frac 1{4\pi
}ln(2\sigma
(\vec{x}\,^{\prime },\vec{x})),
\end{equation}
where $2\sigma (\vec{x}\,^{\prime },\vec{x})$ is the geodesic
distance between the two points $\vec{x}$ e $\vec{x}\,^{\prime }$. Now,
we
expand the regularized Green function in terms of $\sigma$, since we are
interested in the  Green function at small values of $\sigma $.
Grats {\it et al.} \cite{grats,grats2,grats3} showed that the 
regularized Green  function in the
coincident limit is given by
\begin{equation}
G_b^{\left( 2\right) }(\vec{r}_0,\vec{r}_0)\left| _{reg}\right.
=\frac{\Omega (
\vec{x})}{4\pi}. \label{greenc}
\end{equation}
This equation shows explicitly that the self-energy appears from the
geometry
of the defect.

\section{The self-interaction of a density of defects}
\subsection{The disclination density}

\begin{figure}
\begin{center}
\rotatebox{-0}{
\resizebox{6cm}{6cm}{\includegraphics{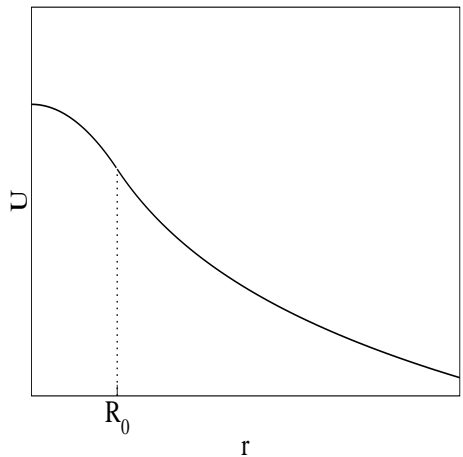}}     }
\caption{The self-energy of a charged particle in the presence of a
disclination density.}
\label{ale1}
\end{center}
\end{figure}

In this subsection we determine the self-energy and the self-force for
a punctual charge in the presence of a distribution of disclinations.
The presence of a density of defects in the continuum medium changes
the field lines of a charged particle making the particle interact with
its own electrical field, resulting in the self-energy of the
particle. Outside the defect, from (\ref{uele}), (\ref{greenc}) and (\ref{cfactordisc}), 
the self-energy is given by
\begin{equation}
U_{out}=-\frac{Q^{2}qR^{2}_{0}}{8\pi}ln(r).
\end{equation}
The behavior of the self-energy is showed in figure \ref{ale1}.
The self-force in the outside region is obtained directly from:
$\vec{F}=-\vec{\nabla}
U$, where $\vec{\nabla} U=e^{(\Omega)}\vec{\nabla_{E}}
U$. Thus, in the outside region,
\begin{equation}
\vec{F}_{out}=\frac{Q^{2}qR_{0}^{2}}{8\pi}\frac{r^{qR_{0}^{2}/2}}{r}\hat{e}_{r}.
\end{equation}
This result is better understood if we introduce a new radial
variable $\rho$ such that
\begin{equation}
\left(1-\frac{qR^{2}_{0}}{2}\right)\rho=r^{\left(1-qR^{2}_{0}/2\right)}.
\end{equation}
In this way, the self-force can be written as
\begin{equation}
\vec{F}_{out}=\frac{Q^{2}(qR^{2}_{0}/2)}{4\pi(1-qR^{2}_{0}/2)}\frac{\hat{e}_{r}}{\rho}.
\end{equation}
We notice that this self-force is equivalent to the force between two
charges of moduli $Q$ and  $Q(qR^{2}_{0}/2)/(1-qR^{2}_{0}/2)$. The
charges are separated by a distance $\rho$.
Inside the defect the particle has the following self-energy
\begin{equation}
U_{in}=\frac{qQ^{2}R^{2}_{0}}{8\pi}\left[\ln(R_{0})+\frac{r^{2}}{2R_{0}^{2}}-\frac{1}{2}\right].
\end{equation}
This shows a harmonic behavior for the electrostatic energy inside
the defect. The self-force is given by
\begin{equation}
\vec{F}_{in}=-\frac{qQ^{2}R_{0}^{2}}{8\pi}\exp\left[\ln(R_{0})+\frac{r^{2}}{2R_{0}^{2}}-\frac{1}{2}\right]r\hat{e}_{r}.
\end{equation}

\begin{figure}
\begin{center}
\rotatebox{-0}{
\resizebox{6cm}{6cm}{\includegraphics{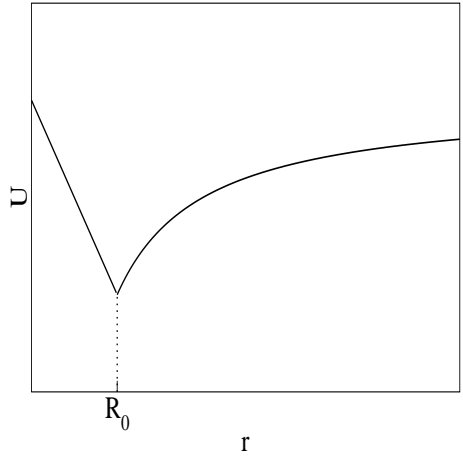}}     }
\caption{The self-energy of a charged particle in the presence of an
edge dislocation density.}
\label{ale2}
\end{center}
\end{figure}
Notice  the discontinuity of the self-force in
the boundary $r=R_{0}$ in spite of  the
self-energy be continuous. Another interesting behavior of the
self-force is that the charge is attracted by the defect density 
independently of the its sign.

\subsection{The edge dislocation density}

In this subsection we determine the self-energy of a punctual charge
in the presence of a density distribution of edge dislocations. 
From (\ref{uele}), (\ref{greenc}) and (\ref{cfactordisl}),The self-energy in the defect region is given by
\begin{equation}
U_{in}=-\frac{Q^{2}\beta r \cos\theta}{8\pi}.
\end{equation}
The behavior of the self-energy, for fixed $\theta$, is shown in 
figure \ref{ale2}. Notice that the self-energy as a function of 
$\theta$ can be either positive or negative, meaning attractive or repulsive 
regions of space. The self-force which acts on the charge
can be easily obtained just by taking the negative gradient
of the energy
\begin{equation}
\vec{F}_{in}=\frac{Q^{2}\beta}{8\pi}\exp(\beta r \cos \theta /4)
\left(-\cos \theta \hat{e}_{r}+\sin \theta \hat{e}_{\theta}\right).
\end{equation}

In the outside region  the self-energy is
\begin{equation}
\label{bardeen}
U_{out}=-\frac{Q^{2}\beta R_{0}^{2} \cos \theta}{8\pi r},
\end{equation}
and the self-force is given by
\begin{equation}
\vec{F}_{out}=\frac{Q^{2}\beta R_{0}^{2}}{8\pi r^{2}}
\exp(\beta R_{0}^{2} \cos \theta /4)
\left(\cos \theta \hat{e}_{r}+\sin \theta \hat{e}_{\theta}\right).
\end{equation}
The expression (\ref{bardeen}) is in agreement with the previous
results obtained by the Potential Deformation Theory~\cite{pr:bs}. 
This approach was used to find the energy interaction 
of electrons and holes in a solid that contains a single edge
dislocation.

\section{Concluding Remarks}

In this work we determined the self-energy of a single charge in the
presence of either a continuous distribution of disclinations or a
continuous distribution of dislocations. We  used the
geometrical theory of defects in solids to describe the defect distributions
in an elastic medium. We also have used  Grats
and Garcia's method for finding the two-dimensional Green function in
a space associated to a conformal metric. This  allowed us to determine
the self-energy and self-force experimented by the charged particle in the
presence of a circular distribution of disclinations. We obtained
the self-energy both for the region inside and outside the distribution.
We notice that the  self-energy is continuous in
the boundary between the internal and external regions but the
self-force is
discontinuous in both cases. The analysis of the behavior of the
self-energy
shows that in the external region the charge ``feels" an 
effective distribution acting as a  single defect in both cases.

Notice that outside the
dislocation density region one has a metric which corresponds to 
an effective dislocation. The self-force due to the distribution
should be compared with the results of reference~\cite{carvalho}, where the 
dipole approximation was used for the metric of a dislocation.
Notice also that the sudden change in the metrics from the defect density
region ($r < R_0$) to the defect-free region ($r>R_{0}$) causes the
curve of the self-energy as function of r to change its concavity.
In the dislocation case, the effect is even more drastic, besides
the change in concavity the derivative (e.g. self-force)
goes discontinuous at $r=R_{0}$. 
Notice also that both in the disclination and in dislocation cases, 
the self-energy is attractive independently of the signal of the charge
$Q$. This gives the possibility of trapping charges in the 
defect density. In the dislocation case there is an additional asymmetry 
in the azimuthal angle, creating regions alternatively of attraction and repulsion 
as one goes around the $z$ axis. 

\noindent
{\bf Acknowledgments}\\
\noindent
We are indebted to Gustavo Camelo for helping with the figures and to 
Mario Henrique de Oliveira for important discussions.
This
work was partially supported by CNPq, FINEP (PRONEX) and CAPES (PROCAD).


\section*{References}
\begin{thebibliography}{99}
%
\bibitem{boo} 
G. E. Volovik, {\it The Universe in a Helium Droplet}
(Oxford University Press)2003.
%
\bibitem{prd:unruh}
W. G. Unruh,
Phys. Rev. D, 51 6(1995).
%
\bibitem{prl:gar} 
L. J. Garay, J. R. Anglin, J. I. Cirac, and P. Zoller
Phys. Rev. lett. {\bf 85}, 4643 (2000).
%
\bibitem{pra:gar}
L. J. Garay, J. R. Anglin, J. I. Cirac and P. Zoller
Phys. Rev. {\bf 63} 023611 (2001).
%
\bibitem{cqg:visser}
Matt Visser,
Class. Quant. Grav. {\bf 15} 1767 (1998).
%
\bibitem{novello}
M. Novello,
Int. J. Mod. Phys A {\bf 17} 29, 4187 (2002).
%
\bibitem{katanaev}
M. Katanaev and I. Volovich, 
 Annals of Physics {\bf 216}, 1(1992).
%
\bibitem{pr:bs}  
J. Bardeen and W. Schockley, Phys. Rev. {\bf  80},  72 (1950).
%
\bibitem{furtado}
C. Furtado and F. Moraes,
Phys Lett. A {\bf 188}, 394(1994)
%
\bibitem{furtado2}
C. Furtado and F. Moraes,
Europhys. Lett.  {\bf 45}, 279(1994).
%
\bibitem{furtado3}
C. Furtado, V. B. Bezerra and F. Moraes,
Europhys. Lett.  {\bf 52}, 1(2000).
%
\bibitem{katanaev2}
M. Katanaev and I. Volovich, 
 Annals of Physics {\bf 272}, 203(1999).
%
\bibitem{dewitt}
C. M. De Witt and B. S. De Witt,
 Physics {\bf 1} 3 (1964)\\
W. G. Unruh,
 Proc. R. Soc. London A{\bf 348}, 447 (1976).
%
\bibitem{vilenkin}
A. Vilenkin,
Phys. Rev. D {\bf 20}, 373 (1979).
%
\bibitem{linet}  
B. Linet, 
Phys. Rev. D {\bf 33}, 1833 (1986).
%
\bibitem{smith}
A. G. Smith,
in Proceedings of Symposium on The Formation and
Evolution of Cosmic String, edited by G. W. Gibbons, S. W. Hawking,
and T. Vachaspati(Cambridge Unversity Press, Cambridge , England, 1990).
%
\bibitem{souradeep}
T. Souradeep and V. Sahni,
 Phys. Rev. D {\bf 46}, 1616 (1992).
%
\bibitem{cqg:furtado} C. Furtado and F. Moraes, Classical Quant. Grav. {\bf 14}, 12, 3425 (1997).
%
\bibitem{guimaraes}
M. E. X. Guimaraes and B. Linet ,
Class. Quantum Grav. {\bf 10}, 1665 (1993).
%
\bibitem{mello1}
E. R. Bezerra, V. B. Bezerra, C. Furtado and F. Moraes,
Phys. Rev. D {\bf 51} 7140 (1994).
%
\bibitem{mello2}
E. R. B. de Mello and C. Furtado,
 Phys. Rev. D {\bf 56} 1345 (1997).
%
\bibitem{carvalho}
A. M. de M. Carvalho, C. Furtado and F. Moraes,
Phys Rev D {\bf 62}, 067504 (2000).
%
\bibitem{azevedo}
S. Azevedo and F. Moraes,
J. Phys. Condens. Matter {\bf 12} 7421 (2000).
%
\bibitem{lorenci}
V. A. de Lorenci and E. S. Moreira Jr,
Phys Rev D {\bf 65}, 085013 (2002).
%
\bibitem{val}
N. R. Khusnutdinov and V. B. Bezerra, Phys Rev D {\bf 64}
083506 (2001)
%
\bibitem{fer} 
F. Moraes, A. M. de M. Carvalho, I.V.L. Costa, F. A. Oliveira
and C. Furtado, 
Phys Rev D {\bf 68} 043512 (2003). 
%
\bibitem{klei}
H. Kleinert, \textit{Gauge fields in Condensed Matter},
Vols. I, II (World Scientific, Singapore, 1989).
%
\bibitem{hol} 
A. Holz, Clas. Quant. Grav. {\bf 5} 1259 (1988).
%
\bibitem{put} 
R. A. Puntigam, Clas.Quant. Grav. {\bf 14}  1129(1997).
%
\bibitem{kol} C. Kohler, Clas. Quant. Grav. {\bf 12}  2977(1995).
%
\bibitem{baush} R. Bausch , R. Schmitz  and L. A. Turski, Phys. Rev.
Lett.{\bf80} 2257 (1998).
%
\bibitem{katanaev3} M. O. Katanaev, Theor.Math.Phys. {\bf 135} 733
(2003).
%
\bibitem{grats}
Yu Grats and A Garcia,
Class. Quantum Gra\-v. {\bf 13}, 189(1996).
%
\bibitem{jackson}
J D Jackson,
{\it Classical Electrodynamics}, Wiley and
Sons, 2$^{nd}$ edition.
%
\bibitem{grats2}
E R B de Mello, V B Bezerra and Yu Grats,
Class. Grav. {\bf 13}, 1915 (1998)
%
\bibitem{grats3}
E R B de Mello, V B Bezerra and Yu Grats,
Modern Physics Lett. A. {\bf 13}, 1427 (1998)
%
\end{thebibliography}
\end{document}